\documentclass[prd,showpcs,amsmath,amssymb,nofootinbib,longbibliography,twocolumn,showpacs,notitlepage,superscriptaddress]{revtex4-1}
\usepackage{multirow}
\usepackage{epsfig}
\usepackage{amsmath}
\usepackage{bm}
\usepackage{times}
\usepackage{graphicx}
\usepackage{color}
\usepackage{slashed}
\usepackage{xcolor}
\usepackage{graphicx}
\usepackage{amsmath}
\usepackage{hyperref}
\usepackage{soul}
\usepackage{lipsum}

\def\bea{\begin{eqnarray}}
\def\eea{\end{eqnarray}}
\def\bean{\begin{equation*}}
\def\eean{\end{equation*}} 

\def\beaal{\begin{align}}
\def\eeaal{\end{align}}
\definecolor{OliveGreen}{rgb}{0,0.6,0}

\begin{document}
\newcommand{\MS}[1]{{\color{green}[MS: #1]}}

\begin{flushright}
KCL-PH-TH-2024-26 
IFT-UAM/CSIC-24-95
\end{flushright}

\title{Detection Prospects of Gravitational Waves from SU(2) Axion Inflation}

\author{Charles Badger}
\email{charles.badger@kcl.ac.uk}
\affiliation{Theoretical Particle Physics and Cosmology Group,  Physics Department, \\ King's College London, University of London, Strand, London WC2R 2LS, United Kingdom}

\author{Hannah Duval}
\email{Hannah.Marie.D.Duval@vub.be}
\affiliation{Theoretische Natuurkunde, Vrije Universiteit Brussel, Pleinlaan 2, B-1050 Brussels, Belgium}

\author{Tomohiro Fujita}
\email{tomofuji@aoni.waseda.jp}
\affiliation{Research Center for the Early Universe, The University of Tokyo, Bunkyo, Tokyo 113-0033, Japan}
\affiliation{Waseda Institute for Advanced Study, Waseda University, 1-6-1 Nishi-Waseda, Shinjuku, Tokyo 169-8050, Japan}

\author{Sachiko Kuroyanagi}
\email{sachiko.kuroyanagi@csic.es}
\affiliation{Instituto de F\'isica Te\'orica UAM-CSIC, Universidad Auton\'oma de Madrid, Cantoblanco, 28049 Madrid, Spain}
\affiliation{Department of Physics, Nagoya University, Nagoya, 464-8602, Japan}

\author{Alba Romero-Rodr\'iguez}
\email{alba.romero-rodriguez@vub.be}
\affiliation{Theoretische Natuurkunde, Vrije Universiteit Brussel, Pleinlaan 2, B-1050 Brussels, Belgium}

\author{Mairi Sakellariadou}
\email{mairi.sakellariadou@kcl.ac.uk}
\affiliation{Theoretical Particle Physics and Cosmology Group,  Physics Department, \\ King's College London, University of London, Strand, London WC2R 2LS, United Kingdom}

\date{\today}

\begin{abstract}
We study detection prospects of a gravitational-wave background (GWB) sourced by SU(2) gauge fields considering all possible observational constraints. More precisely, we consider bounds set by cosmic microwave background measurements, primordial black hole overproduction, as well as backreaction of the gauge fields on the background evolution. Gravitational-waves data from the first three observing runs of the LIGO-Virgo-KAGRA Collaboration show no evidence for a GWB contribution from axion inflation. However, we are able to place conservative constraints on the parameters of the SU(2) inflation with current data. We investigate conditions on the inflationary potential that would lead to a detectable signal that evades astrophysical and cosmological constraints and discuss detection prospects for third generation networks. 
\end{abstract}
\vspace{-2mm}

\maketitle

%===============================================================%
\section{Introduction}
\label{Sec:Intro}
%===============================================================%

The Universe is expected to be permeated by a gravitational-wave background (GWB) resulting from gravitational waves (GW) of cosmological origin, along with the superposition of GWs from weak and distant astrophysical sources. Examples of cosmological GWs include those generated by first-order phase transitions, topological defects, or an inflationary era (see, for example, ~\cite{Kalogera:2021bya,LISACosmologyWorkingGroup:2022jok} for a comprehensive overview). GWB searches conducted by interferometer experiments, such as those carried out by the LIGO-Virgo-KAGRA (LVK) collaboration~\cite{LIGOScientific:2016jlg, LIGOScientific:2019vic,KAGRA:2021kbb}, represent a unique and valuable tool for imposing constraints on particle physics models beyond the Standard Model at energy scales exceeding those presently achievable by particle accelerators.

Inflation models generally predict a primordial GWB arising from quantum fluctuations in spacetime stretched across the horizon during the accelerated expansion epoch. While minimal slow-roll inflation predicts a nearly flat GWB spectrum with an amplitude typically too small for detection by current and near-future experiments, considering additional degrees of freedom or new symmetry patterns during inflation may lead to a blue-tilted GW spectrum~\cite{Bartolo:2016ami}, making it an interesting case for GW interferometers.

In this context, axion inflation has garnered significant attention for its capacity to produce strong GW signals detectable by interferometer experiments. In this model, the flat direction responsible for driving inflation is characterized by a pseudoscalar axion field, which couples to a gauge field~\cite{Anber:2009ua,Cook:2011hg,Dimastrogiovanni_2013}. As a consequence of the gauge field background, this model effectively becomes multi-field, giving rise to several interesting prospects for cosmological observations~\cite{Barnaby:2011qe,Barnaby:2011vw} such as features in the cosmic microwave background (CMB) measurements~\cite{Barnaby:2010vf,Meerburg:2012id}, formation of primordial black holes (PBHs)~\cite{Domcke:2017fix}, and generation of the chiral GWB~\cite{Cook:2011hg,Garcia-Bellido:2016dkw,Domcke:2016bkh,Dimastrogiovanni:2016fuu,Garcia-Bellido:2023ser}. 

%Let us consider 
Substantial work has been done in the context of the U(1) gauge field coupled to the inflaton field, while we can also consider a particle physics model in which the inflaton field interacts with non-abelian SU(2) gauge fields that have an isotropic vacuum expectation value (VEV).
%as their initial state. 
Slow-roll inflation is realized through the transfer of axionic energy into classical gauge fields, %dubbed as
known as as chromo-natural inflation~\cite{Adshead:2012kp,Dimastrogiovanni:2012st}.
The presence of background gauge fields with non-zero VEV induces linear couplings between metric and gauge field tensor perturbations, resulting in the generation of a significant GWB~\cite{Maleknejad:2016qjz,Thorne:2017jft}. However, certain models, such as the originally proposed cosine inflaton potential, have already been ruled out by CMB observations~\cite{Adshead:2013qp,Adshead:2013nka}. Nevertheless, extensive studies are underway to explore the possibility of evading the constraints at the CMB scale whilst still producing interesting signals at scales relevant for interferometric studies. Examples include employing different inflation potentials~\cite{Obata:2016xcr,Caldwell:2017chz,DallAgata:2018ybl}, exploring realizations where the axion field is not the inflaton~\cite{Dimastrogiovanni:2016fuu,McDonough:2018xzh}, introducing a spontaneously broken gauge symmetry~\cite{Adshead:2016omu}, or considering a delayed emergence of the chromo-natural system~\cite{Obata:2014loa,Obata:2016tmo,EmergingDomcke,Fujita:2022jkc}. 

In what follows, we investigate detection prospects of the GWB originating from axion-SU(2) gauge field inflation. Our objective is to study illustrative models that can be tested by forthcoming GWB searches while carefully accounting for all cosmological constraints on the model imposed by CMB measurements, PBH overproduction, as well as backreaction of the gauge fields on the background evolution~\cite{Papageorgiou:2019ecb,Iarygina:2023mtj}. We particularly emphasize the relevance of these models to ground-based interferometer experiments, including the LVK detector network, as well as third-generation (3g) detector networks like the Einstein Telescope (ET)~\cite{Maggiore_2020,Punturo:2010zz} and the Cosmic Explorer (CE)~\cite{Evans:2021gyd}. To achieve this,  
we first explore the parameter space in model independent way. Then we investigate two concrete inflation models. First, we consider two-stage inflation where the CMB and interferometer scales are governed by different potential to satisfy observational constraints. 
Subsequently, we examine Starobinsky model~\cite{Starobinsky:1980te} as another representative example, known for their agreement with CMB observations. In our study, we focus on the SU(2) gauge group, though our work can be easily extended to other groups $\cal G$ by properly embedding SU(2) into $\cal G$ \cite{Fujita:2021eue}. 

Furthermore, we conduct
the first Bayesian inference search using the latest publicly available data from the first three LVK observing runs to constrain parameters of the SU(2) gauge field inflation.
We also simulate future GWB searches, assuming Advanced LIGO A+ \cite{Aplusdesign}, as well as a network of 3rd generation detectors. This network comprises two ET-like detectors and two CE-like detectors \cite{Hild_2011, borhanian2024listening}. These simulations demonstrate the increased constraining power of future detectors.

This paper is organized as follows. In Sec.~\ref{Sec:GW}, we review GWB production in the context of SU(2) chromo-natural inflation. We present the observational constraints from CMB and PBH overproduction and discuss prospects of GWB detectability, without specifying a particular inflationary potential. 
In Sec.~\ref{sec: models}, we examine two specific inflation models. One is inflation with a piecewise linear toy model potential, which serves as a working example. The other model is the well-studied Starobinsky potential, where we explicitly illustrate how it leads to a GWB that is effectively undetectable by a 3g detector network. 
In Sec.~\ref{sec: PE}, we  perform a Bayesian inference search to constrain parameters of the simple toy model we have presented previously, using LVK data. We subsequently discuss prospects for Advanced LVK A+ and 3g detectors. We present our conclusions in Sec.~\ref{sec: conclusions}.

%===============================================================%
\section{Axion-SU(2) Gauge Field Inflation}
\label{Sec:GW}
\subsection{Formalism}
%===============================================================%

The action for chromo-natural inflation (CNI) reads~\cite{Adshead:2012kp, Dimastrogiovanni_2013}

\begin{equation}
\begin{split}
    S_{\rm CNI} = \int d^4x \sqrt{-{\rm det}(g_{\mu\nu})} \Big[\frac{M_{\rm Pl}^2}{2}R - \frac{1}{2}(\partial\phi)^2 \\
    - V(\phi) -\frac{1}{4}F_{\mu\nu}^a F^{a\mu\nu} + \frac{\alpha_f}{4}\phi F_{\mu\nu}^a \tilde{F}^{a\mu\nu}\Big] \,,
    \label{eq: CNI_Lag}
\end{split}
\end{equation}
where $M_{\rm Pl}$ denotes the reduced Planck mass, $R$ stands for the Ricci scalar, $\phi$ is a pseudo-scalar inflaton field (axion) with potential $V(\phi)$, $F_{\mu\nu}^a = \partial_\mu A_\nu^a - \partial_\nu A_\mu^a - g\varepsilon^{abc}A_\mu^b A_\nu^c$ is the field strength of an SU(2) gauge field $A_\mu^a$, $\tilde{F}^{a\mu\nu} = \varepsilon^{\mu\nu\rho\sigma} F_{\rho\sigma}^a /(2\sqrt{-{\rm det}(g_{\mu\nu})})$ is its dual, $\alpha_f$ denotes the coupling constant between the axion and the gauge field, and $g$ stands for the gauge coupling constant.

We use an isotropic ansantz for the homogeneous part of the gauge field, namely
\begin{equation}
    A^a_0 = 0, \quad A^a_i = \delta_i^a a(t) Q(t),~
    \label{eq: fieldAmsatz}
\end{equation}
which is an attractor solution during inflation~\cite{Maleknejad:2013npa, Wolfson:2021fya,Murata:2022qzz}. The Friedmann equation reads
\begin{equation}
    3M_{\rm Pl}^2 H^2 = \frac{\dot{\phi}^2}{2} + V(\phi) + \frac{3}{2}(\dot{Q} + HQ)^2 + \frac{3}{2}g^2 Q^4~,
    \label{eq: EinsteinEq}
\end{equation}
where $H\equiv \dot a/a$ is the Hubble parameter and the dot stands for a cosmic time derivative $(\dot\phi \equiv \partial \phi/ \partial t)$. The background equations of motion for the inflaton and gauge fields are ~\cite{Adshead:2012kp, Dimastrogiovanni_2013}
\begin{equation}
    \ddot{\phi} + 3H\dot{\phi} + V'(\phi) = -3g\alpha_f Q^2 (\dot{Q} +HQ)~,
    \label{eq: phi_EOM}
\end{equation}
\begin{equation}
    \ddot{Q} + 3H\dot{Q} + (\dot{H} + 2H^2)Q + 2g^2Q^3 = g\alpha_f \dot{\phi} Q^2~,
    \label{eq: Q_EOM}
\end{equation}
respectively, where $V'(\phi) = \partial V(\phi) / \partial \phi$.

Solving Eqs.~\eqref{eq: EinsteinEq}-\eqref{eq: Q_EOM}, one finds that the gauge field gets a non-vanishing VEV, equal to $Q\simeq (-V'/3\alpha_f gH)^{1/3}$ during inflation~\cite{Adshead:2012kp}. Since the gauge field provides additional effective friction to the inflaton through Eq.~\eqref{eq: phi_EOM},  slow-roll inflation can be more easilly achieved.

The non-Abelian gauge field $\delta A_\mu^a$ makes significant contributions to perturbations that can be decomposed into scalar, vector, and tensor components around the homogeneous solution~\eqref{eq: fieldAmsatz}. Only the tensorial component is amplified by tachyonic instability~\cite{Dimastrogiovanni_2013, Adshead:2013nka} that at linear order sources GWs. The GW background is described in terms of its energy density spectrum via:
\begin{equation}
    \Omega_{\rm{GW}}(f) = \frac{1}{\rho_c}\frac{d\rho_{\rm{GW}}}{d{\rm{log}}~f}
    \label{eq: GW_Gen}
\end{equation}
where $\rho_c = 3c^2 H_0^2 / (8\pi G)$ is the critical energy density of the Universe.
The sourced contribution to the primordial GWs has been analytically estimated as~\cite{EmergingDomcke}
\begin{equation}
    \Omega_{\rm GW}^{\rm (SU2)}(k) \simeq \frac{\Omega_{R,0}}{24} \Big(\frac{\xi^3 H}{\pi M_{\rm Pl}}\Big)^2_{\xi = \xi_{cr}} \Big(\frac{2^{7/4} H}{g\sqrt{\xi}} e^{(2-\sqrt{2})\pi\xi}\Big)^2_{\xi = \xi_{\rm{ref}}},
    \label{eq: GW_SU2}
\end{equation}
where $\Omega_{R,0} = 8.4 \times 10^{-5}$ and $\xi$ is a non-trivial function of the inflaton velocity parameter,
\begin{equation}
    \xi = \frac{\alpha_f\dot{\phi}}{2H}\,.
    % = \frac{\alpha}{2}\phi_N\,.
    \label{eq: xi_def}
\end{equation}
%The kinetic energy of the inflaton is transferred into the gauge sector in this class of models.
We note that the first parenthesis in Eq.~\eqref{eq: GW_SU2} uses $\xi_{\rm{cr}} = \xi ~(x=1)$, while the second parenthesis uses $\xi_{\rm{ref}} = \xi ~(x=(2+\sqrt{2})\xi_{\rm{cr}})$ with $x=-k\tau$ for conformal time $\tau$. This gauge contribution is applicable in the non-Abelian regime, conservatively defined as~\cite{EmergingDomcke}
\begin{equation}
    0.008e^{2.8\xi} 
    %\ll 
    \gtrsim
    1/g~.
    \label{eq: nonAbelReg}
\end{equation}

The total GW spectrum 
also includes the standard vacuum contribution
\begin{equation}
    \Omega_{\rm GW}^{\rm (vac)}(k) = \frac{\Omega_{R,0}}{12\pi^2}\frac{H^2}{M_{\rm Pl}^2}\,.
    \label{eq: GW_vac}
\end{equation}
One may wonder whether the sourced tensor perturbations would produce a signal detected by GW interferometers. In fact,  Ref.~\cite{EmergingDomcke} reported that
the sourced GWs on the LVK frequency range can be greatly enhanced so that $\Omega_{\rm GW}^{\rm (SU2)}\gg \Omega_{\rm GW}^{\rm (vac)}$. Although the sourced contribution does not reach the LVK sensitivity for their specific parameters, they stressed that a different parameter choice could change the result. 

In practice, GW spectra are typically studied over e-folds $N$, but can be studied in frequency using~\cite{Bartolo:2016ami}
\begin{equation}
\begin{split}
    N = N_{\rm{CMB}} + \log\Big (\frac{k_{\rm{CMB}}}{0.002~\rm{Mpc}^{-1}} \Big ) - 40.3 \\ - \log\Big(\frac{f}{\rm{Hz}}\Big) + \log\Big(\frac{H(N)}{H_{\rm{CMB}}}\Big) \,,
    \label{eq: Ntof}
\end{split}
\end{equation}
where $H_{\rm{CMB}}$ is the Hubble expansion rate at the CMB scale, assumed to be $N_{\rm{CMB}}$ efolds before the end of inflation and we assume $k_{\rm{CMB}} = 0.002~\rm{Mpc}^{-1}$. %\HD{explain $N_{\rm{CBC}}$}.

%===============================================================%
\subsection{Observational Constraints}
\label{Sec:Constraints}
%===============================================================%

Even though the SU(2) gauge field can enhance GWs during inflation, there are certain limitations. We  summarize below the relevant observational constraints.

%===============================================================%
\subsubsection{Cosmic Microwave Background }
%===============================================================%
The energy scale of inflation corresponds to the amplitude of the tensor perturbation spectrum $\Omega_{\rm GW}^{\rm (vac)}$. 
The tensor-to-scalar ratio $r$ is related to the Hubble expansion rate during inflation evaluated at the CMB scale, $H_{\rm CMB}$:
\begin{equation}
    H_{\rm CMB} = 2.7 \times 10^{14} r^{1/2} {\rm GeV} \,.
\end{equation}
The latest upper bound on $r$, obtained by the joint analysis of the BICEP2, Keck Array and BICEP3 CMB polarization experiments, is $r<0.036$ at $95\%$ CL~\cite{BICEP:2021xfz}. It implies an upper bound on the inflation scale: $H_{\rm CMB} < 5.1\times 10^{13}$GeV $= 2.1\times 10^{-5}~\rm{M_{Pl}}$. 
This bound is valid under the assumption that the SU(2) contribution is negligible at CMB scales.

\subsubsection{PBH overproduction and backreaction effects}

The amplified tensorial part of the gauge field perturbation results in the enhancement of primordial curvature fluctuations 
at the second order in cosmological perturbation. 
Similar to GWs, the dimensionless power spectrum of curvature fluctuations has two contributions
\begin{equation}
    \Delta_s^2(k)\equiv \frac{k^3}{2\pi^2}P_\zeta
    = \left(\frac{H^2}{2\pi\dot{\phi}}\right)^2
    + (\Delta_s^2)^{\rm (SU2)}\,.
\label{eq: deltaS2}
\end{equation}
The first term on the right-hand side of the equation above represents the standard contribution from the linear inflaton fluctuations, whereas the second term accounts for the contribution from SU(2) gauge fields, which may dominate $\Delta_s^2$ at small scales;
its numerical fit is given in Ref.~\cite{EmergingDomcke}.

PBHs are produced when such large curvature fluctuations re-enter the horizon during the radiation-dominated epoch after inflation. These PBHs must satisfy constraints on their abundance, derived from cosmological and astrophysical data~\cite{Carr:2009jm, Carr:2020gox}. 
When curvature fluctuations are non-linearly sourced, they are given by
the convolution of two Gaussian modes, hence curvature perturbations obey $\chi^2$ statistics. As discussed in \cite{Linde:2012bt,Garcia-Bellido:2016dkw}, such a non-Gaussian distribution predicts a PBH abundance larger than the corresponding one for a Gaussian distribution. Therefore,~\cite{Garcia-Bellido:2016dkw}
\begin{equation}
\Delta_s^2 \lesssim \left\{
\begin{array}{cc}
   10^{-4}  &  (\chi^2\ {\rm case} ) \\
   10^{-2}  &  ({\rm Gaussian \ case} )
\end{array}\right.\,,
\end{equation}
The scenario we are studying falls in the case of $\chi^2$ statistics, and we will show that the PBH bound excludes a large part of the parameter space.

Let us  briefly discuss a possible caveat of the PBH bound for the $\chi^2$ statistics.
Recent lattice simulations of axion inflation with a U(1) gauge field have 
reported that the distribution of curvature perturbations generated in a strong backreaction regime, in which the produced gauge field fluctuations affect the background inflationary dynamics, approaches a Gaussian distribution~\cite{Caravano:2022epk}. The regime where large GWs are produced is indeed likely to experience a strong backreaction
and hence it is important to estimate the corresponding parameter space.

The backreaction from the amplified gauge field perturbations to the background dynamics appears as a new term $\mathcal{T}_{\rm BR}^Q$ in Eq.~\eqref{eq: Q_EOM}~\cite{Papageorgiou:2019ecb}.
To quantify the strength of the backreaction, we introduce a new quantity $\kappa$ defined as 
\begin{equation}
    \kappa \equiv \frac{g\alpha_f \dot{\phi} Q^2}{\mathcal{T}_{\rm BR}^Q}
    \simeq g \bigg (\frac{24\pi^2}{2.3e^{3.9 m_Q}} \frac{m_Q^2}{1+m_Q^2} \bigg )^{-1/2},
    \label{eq: kappa}
\end{equation}
where $m_Q\equiv gQ/H$ and we used the slow-roll solution for $Q$ and the numerical fit for $\mathcal{T}_{\rm BR}^Q$~\cite{Papageorgiou:2019ecb}. 
We expect that if the PBH bound could be 
relaxed
by the gaussianization effect due to the backreaction, it should happen at around $\kappa=1$. 

In the strong backreaction regime, $\kappa > 1$, the aforementioned expressions are no longer valid and one needs dedicated analysis,
which is beyond the scope of our present study.

%===============================================================%
\subsection{Gravitational-wave background detectability}
\label{Sec:GenResults}

Let us study prospects of detecting GWs from the SU(2) gauge field taking into account all observational constraints discussed above. Determining the resulting background without knowledge of the inflaton potential is difficult, thus we 
make some simplifications. 
We assume that the resulting $\xi$ over the GW ground-based detectors frequency range is approximately constant. Since we expect $\xi$ to grow over increasing frequencies, this assumption gives a constant, lower bound on the $\Omega_{\rm{GW}}$ in the aforementioned frequency range.

\begin{figure}[h!]
    \includegraphics[width=0.49 \textwidth]{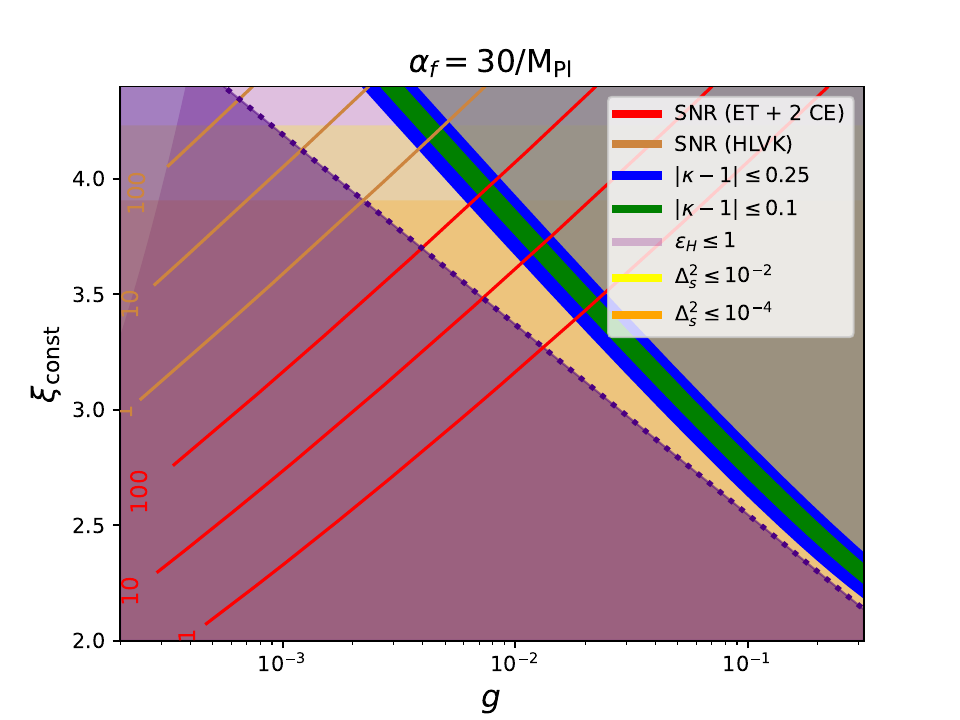}
    \includegraphics[width=0.49 \textwidth]{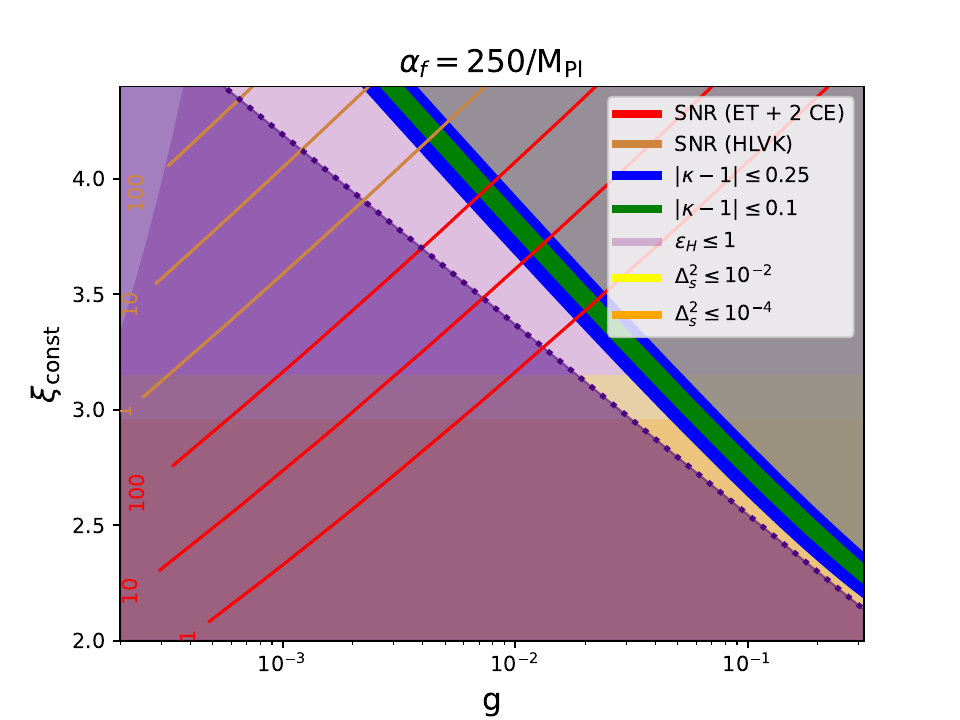}
    \caption{
    Backreaction and SNR prospects as a function of 
    the gauge coupling constant $g$ and the velocity parameter $\xi(f) = \xi_{\rm{const}}$ for two values of the coupling constant, $\alpha_f = 30/M_{\rm Pl}$ (top) and $250/M_{\rm Pl}$ (bottom). The Hubble rate on the CMB scale is assumed to be $H_{\rm{CMB}} = 2.1 \times 10^{-5}$ $M_{\rm Pl}$. The brown and red lines are the SNR contours calculated assuming 1 year of observation 
    %of the LVK and ET + 2 CE network
    with the design LVK and ET + 2 CE network  sensitivities, respectively. The blue and green bands represent $|\kappa -1|\leq 0.25$ and $0.1$, where we expect that the PBH overproduction bound could be relaxed by gaussianization. The light purple region represents the slow-roll parameter satisfying $\epsilon_H \leq 1$, where inflation is still ongoing. Yellow and orange regions correspond to the parameter space where the primordial curvature spectrum stays below the PBH upper limit assuming Gaussian ($\Delta_s^2 \leq 10^{-2}$) and $\chi^2$ statistics ($\Delta_s^2 \leq 10^{-4}$). Additionally, we shade the Abelian regime in magenta (below the black dotted line) and the strong backreaction regime in gray (upper region of the $|\kappa -1|$ band), in both of which the estimation of the SNR is not valid and is beyond the scope of our study.}
    \label{fig: GeneralParam_H_range}
\end{figure}

We plot the observationally valid parameter space described in Sec. ~\ref{Sec:Constraints} with the signal-to-noise ratio (SNR) of the simplified resulting $\Omega_{\rm{GW}}$ in the current LVK~\cite{buikema2020sensitivity} and design ET + 2 CE network in Fig.~\ref{fig: GeneralParam_H_range}, for two values of the coupling constant $\alpha_f = 30/M_{\rm Pl}$ and $250/M_{\rm Pl}$. For an ET + 2 CE network, 
we assume a triangular ET detector configuration located at the Virgo detector site, and two L-shaped CE detectors located at the Hanford and Livingston detector sites, respectively. We additionally plot regimes where inflation is still ongoing ($\epsilon_H \leq 1$) and that produces spectra below the PBH upper limit assuming Gaussian ($\Delta_s^2 \leq 10^{-2}$) and $\chi^2$ statistics ($\Delta_s^2 \leq 10^{-4}$).

It is worth noting that the GWB amplitude in Eq.~\eqref{eq: GW_SU2}, which we have used to calculate the SNR, is valid only in the non-Abelian regime. Additionally, the dynamics would be significantly altered in the strong backreaction regime, rendering the GWB estimation invalid. Thus, our parameter space exploration is constrained in the region between $|\kappa - 1|$ band and the black dotted line which is the conservative Abelian/non-Abelian boundary given by Eq.~\eqref{eq: nonAbelReg}.

For  $\alpha_f = 250/M_{\rm Pl}$, the parameter space that avoids PBH overproduction $\Delta_s^2 \leq 10^{-2}$ leads to a $\rm{SNR} < 1$ even for an ET + 2 CE network. 
Increasing $\alpha_f$ strengthens scalar perturbation $\Delta_s^2$,  Eq.~\eqref{eq: deltaS2}, thus further restricting the portion of the parameter space below the PBH bound.

%However, 
In more realistic models, $\xi$ evolves over frequency.  Our result implies that $\xi$ must remain within a narrow band between the Abelian and strong backreaction regimes 
%to remain physically allowed. 
to expect a large GWB sourced during the non-Abelian regime and a reliable GWB prediction without considering backreaction effects.
For example, assuming $H_{\rm{CMB}} = 2.5 \times 10^{-5} ~M_{\rm Pl}$ and $g = 10^{-2}$ one would need $3.35 \leq \xi(5-5000 ~\rm{Hz}) \leq 3.52$ - a relatively tight range given that the inflaton velocity can increase significantly. 
This poses a challenge in finding an inflationary model that satisfies such conditions, where $\xi$ increases but and remains within the allowed region in the ground-based detectors' frequency range.

\section{Models}
\label{sec: models}

While previously we have discussed GWB detectability in a model-independent framework, in the following we analyse a toy model designed in such a way that the produced GWB is detectable. Notably, this model predicts constant velocity parameter $\xi$.
%and we consider a two-stage inflation so that prediction for the interferometer scale can be made analytically. 
We subsequently investigate the well-studied Starobinsky potential. 
%In this case, we study more realistic spectra evolution by numerically solving the field evolution from the CMB to interferometer scales.

\subsection{Toy Model}
\label{sec: toy_results}

We consider
a piecewise linear toy model potential ~\cite{Martin_2012,Starobinsky:1992ts}
\begin{equation}
    V(\phi)=\left\{
    \begin{array}{rl}
        V_0 + A_+(\phi - \phi_0), & \text{for }\phi > \phi_0\\
        V_0 + A_-(\phi - \phi_0), & \text{for }\phi < \phi_0
    \end{array}\right.
\end{equation}
where $V_0$ is the energy scale of the potential, and $A_+$, $A_-$ define the different potential slopes over $\phi$, changing slope at $\phi(N=N_0) = \phi_0$ and finishing at value $\phi(N=0) = \phi_{\rm{end}}$ where $N_0$ %corresponds 
is the number of e-folds corresponding
to the reference frequency $f_0$. 
Note that $f_0$ denotes also the frequency where the spectrum moves from the Abelian to the non-Abelian regime.
We assume that it lies between CMB and interferometer scales, so that we have an Abelian regime at the CMB scale and a non-Abelian regime at the interferometer scale. This is the easiest way to satisfy the observational constraints at the CMB scale.

Within the slow-roll approximation one can 
show that the resulting inflaton velocity is approximately constant 
\begin{equation}
    \xi=\left\{
    \begin{array}{rl}
        \xi_{\rm{CMB}} = A_+\alpha_f / 2V_0, & \text{for }\phi > \phi_0\\
        \xi_0 = A_-\alpha_f / 2V_0, & \text{for }\phi < \phi_0
    \end{array}\right.
\end{equation}
where the initial inflaton velocity $\xi(N=N_{\rm{CMB}}) = \xi_{\rm{CMB}}$ must remain below the initial inflaton velocity observed by CMB experiments of $\xi_{\rm{CMB}} < 2.5$ at 95\% CL~\cite{Planck:2019kim}. More robust evaluations of this model's behavior can be found in ~\cite{Martin_2012}. 
Thanks to this constant velocity behaviour,
the prospects of this model match those described in Fig.~\ref{fig: GeneralParam_H_range}, with $\xi_0 = \xi_{\rm{const}}$, in which we find a parameter space testable by interferometer experiments.

\subsection{Starobinsky Potential}
\label{sec: Staro_Results}

We additionally study the prospects of detecting gravitational spectra from the Starobinksy (or $R^2$) potential~\cite{Starobinsky:1980te}
\begin{equation}
    V(\phi) = \mathcal{R}^4 [1-e^{-\gamma\phi}]^2,
    \label{eq: R2_Def}
\end{equation}
where $\gamma = \sqrt{2/3}$ and $\mathcal{R}$ is an energy scale of the potential.

\begin{figure}
    \centering
    \includegraphics[width=0.5\textwidth]{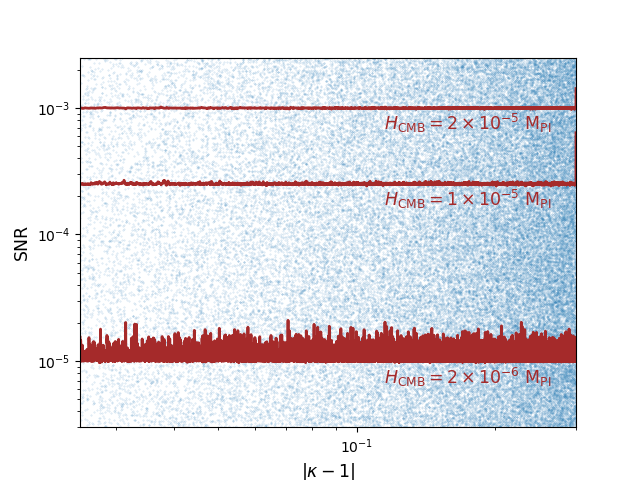}
    \caption{
    The values of the SNR calculated for a detector network of 1 ET + 2 CE interferometers assuming 1 year of observation, shown with the backreaction parameter $|\kappa - 1|$ in the horizontal axis. Each point was randomly sampled by changing the model parameters ($\alpha_f$,  $N_{\rm{CMB}}$, $g$, $H_{\rm{CMB}}$). The red lines are contours of the Hubble rate at the CMB scale, $H_{\rm CMB} = 2\times 10^{-6}~M_{\rm Pl}$, $10^{-5}~M_{\rm Pl}$, $2\times 10^{-5}~M_{\rm Pl}$, indicating that $H_{\rm{CMB}}$ is the majour factor in determining the SNR. }
    \label{fig: KappavsSNR_Staro}
\end{figure}

To investigate whether this model can lead to detectable GWB, we numerically solve the field evolution during inflation. In Fig.~\ref{fig: KappavsSNR_Staro}, we take 160,000 random uniform samples ranging from $0 \leq \alpha_f \leq 40/\rm{M_{\rm Pl}}$, $50 \leq N_{\rm{CMB}} \leq 60$, $-3 \leq \log_{10} g \leq 0$ and $-6 \leq \log_{10} H_{\rm{CMB}} \leq -4.5$ and calculate the resulting $\kappa$ and SNR for an ET + 2 CE network assuming 1 year of observation. 
We find that the assumed Hubble value at CMB $H_{\rm{CMB}}$ plays 
the most significant role in determining the SNR
and that $N_{\rm{CMB}} \geq 55$ and $\alpha_f \leq 24/\rm{M_{\rm Pl}}$ combinations result in $|\kappa - 1| \leq 0.1$.  
However, all simulated spectra with backreaction ratio in this range result in $\rm{SNR} \leq 2.5 \times 10^{-3}$, implying that a suitable Starobinksy potential model is effectively undetectable in a network composed by ET + 2 CE interferometers.

\subsection{Example gravitational and scalar perturbation spectra}

\begin{figure}
    \includegraphics[width=0.49\textwidth]{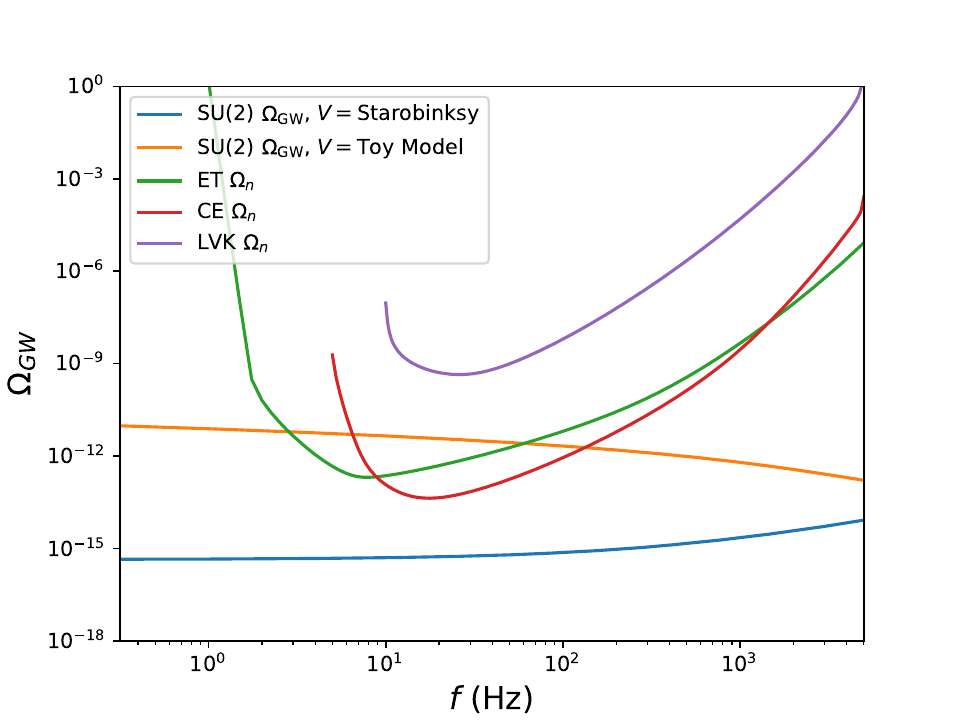}
    \caption{GWB spectra %in units of 
    $\Omega_{\rm GW}$ for the toy model described in Section \ref{sec: toy_results} (orange) and the Starobinsky model (blue), described in Section \ref{sec: Staro_Results}. 
    LVK, ET, and CE sensitivities are also shown.}
    \label{fig: Model_GW_examples}
    \includegraphics[width=0.49\textwidth]{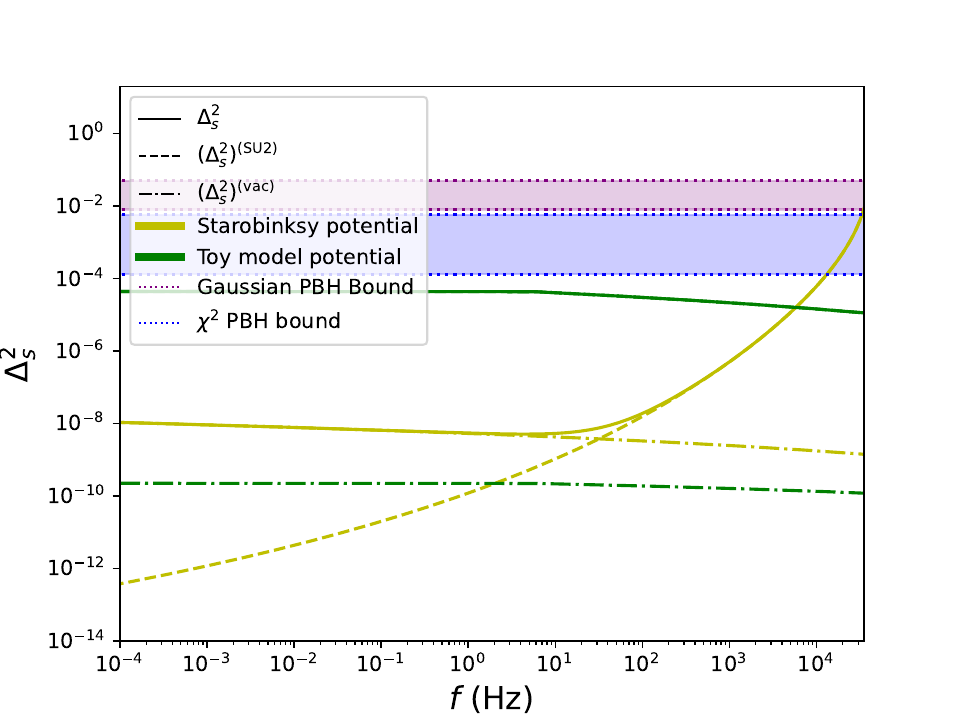}
    \caption{Scalar density perturbation $\Delta_s^2$ 
    plotted as a function of the corresponding GW frequency.
    For both the Starobinksy and the toy model, the vacuum contribution is plotted in a dashed line and the SU(2) contribution is plotted in a dashed-dotted line, while their total contribution in a solid one. The PBH upper bound assuming Gaussian (purple) and $\chi^2$ (blue) statistics are also plotted. }
    \label{fig: Model_deltaS_examples}
\end{figure}

Assuming a toy model with $\xi_{\rm{CMB}} = 0.02$, $\xi_0 = 3.77$, $g=5.32\times 10^{-3}$, $H_{\rm{CMB}} = 2.15\times 10^{-5}$, and the largest, CMB compatible scenario from the Starobinsky model,
we plot the resulting GWs and scalar perturbation spectra in Figs.~\ref{fig: Model_GW_examples} and \ref{fig: Model_deltaS_examples}, respectively. Note that these choices are compatible with CMB observations, described in Sec.~\ref{Sec:Constraints}. One can see that, for the toy model, the resulting $\Omega_{\rm{GW}}$ is above ET, CE sensitivity, and $\Delta_s^2$ is below even the most conservative PBH bound. 

\section{Parameter Estimation}
\label{sec: PE}

Since the toy model we have previously discusses has promising prospects, we analyse in the following how well such a model can be constrained with ground-based detectors.
We first perform a parameter estimation search and set constraints using data from the first three LVK observing runs, and then discuss prospects with advanced LIGO and third-generation detector networks using simulated data. 

\subsection{Methodology}
\label{sec: methodology}

We adopt the conventional LVK stochastic hybrid approach, which combines frequentist and Bayesian analysis techniques~\cite{PhysRevLett.109.171102}. Specifically, we use certain frequentist statistics, namely $\hat C_{IJ}(f)$ and $\sigma_{IJ}(f)$, combined over the entire assumed observation period. These frequentist statistics represent the cross-correlation estimator of the GWB as a function of frequency, obtained from data collected by detectors I and J, with $\sigma^2_{IJ}(f)$ denoting its variance~\cite{PhysRevD.59.102001}. 

Subsequently, these spectra serve as the input data for a Bayesian analysis~\cite{PhysRevLett.109.171102, PhysRevX.7.041058, PhysRevD.102.102005} aimed at computing posterior probability distributions of the axion inflation model parameters. We employ the following Gaussian log-likelihood for a single detector pair~\cite{PhysRevLett.109.171102}
\begin{align}
    \log p(\hat C_{IJ}(f) | \boldsymbol{\theta}_{\rm gw}, \lambda) \propto -\frac{1}{2}\sum_{f}\frac{\left[\hat C_{IJ}(f) - \lambda \, \Omega_{\rm gw}(f, \boldsymbol{\theta}_{\rm gw}) \right]^2}{\sigma_{IJ}^2(f)},
    \label{eq:pe_ms:likelihood}
\end{align}
where $\Omega_{\rm gw}(f, \boldsymbol{\theta}_{\rm gw})$ is the assumed model describing the GWB, dependent on a set of parameters $\boldsymbol{\theta}_{\rm gw}$. The parameter $\lambda$ captures calibration uncertainties of the detectors~\cite{Sun:2020wke} and is marginalized over~\cite{Whelan:2012ur}. In order to perform a multi-baseline study, we add all log-likelihoods of the individual baselines.

\subsection{Constraints on the toy model using LVK GWB search data}\label{sec: PE_O3}

Here, we perform a Bayesian analysis search, as described in Section~\ref{sec: methodology} to constrain the parameters characterising the toy model, described in Sec.~\ref{sec: toy_results}. 
The results presented below are produced using the Python packages {\tt pygwb}~\cite{Renzini:2023qtj} and {\tt Bilby}~\cite{Ashton:2018jfp} employing the {\tt Dynesty} sampler. 

The relevant parameters for the toy model are $\boldsymbol{\theta}_{\rm gw} = (N_{\rm CMB}, f_{0}, \phi_{\rm end}, A_{+}, A_{-}, V_{0}, \alpha_{\rm f}, g)$. The priors for these parameters are given in Table~\ref{table:priors}. 
It is typical to assume inflation in the slow-roll regime to last between 50 to 60 e-folds, and we assume the reference frequency $f_0$ to be less than 10~Hz, otherwise we transition to the non-Abelian regime during or after the ground detector frequency regime. We expect $V_0 \sim V(N = N_{\rm{CMB}})$ with Planck upper bound $1.875 \time 10^{-9}~\rm{Mpl}^4$ ~\cite{Planck:2018jri}, and $A_+, A_- \ll V_0$, thus we pick a broad range to allow for an independent search. Fig.~\ref{fig: GeneralParam_H_range} proved that detection prospects weaken when assumed $\alpha_f$ is too large, thus a generous upper limit of 250 is chosen. Similarly, one can see from the same figure that too small of a $g$ may fall outside the $\epsilon_H \leq 1$ regime, thus we choose a lower bound of $10^{-5}$ out of good caution. To remain physical, we must have $\phi_{\rm{end}} >0$, thus 0 is chosen for a lower band.

In both cases the resulting posteriors provides bounds on the toy model's potential. However, it is of interest to evaluate how well cosmological parameters can be constrained from such a search. One can transform the posteriors of 
\{$A_+$, $A_-$, $V_0$\}
to \{$H_{\rm{CMB}},\xi_{\rm{CMB}}, \xi_0$\} using the work from ~\cite{Martin_2012}. 
For all Bayesian inference searches,  
we show both cases where we assume
Gaussian ($\Delta_s^2 \leq 10^{-2}$) and $\chi^2$ ($\Delta_s^2 \leq 10^{-4}$) statistics for the PBH bound\footnote{Placing such bounds can be done with bilby function \texttt{priors.Constraint}. This slices the prior parameter space to fit specified conditions - in our case: $\Delta_s^2(\boldsymbol{\theta}_{\rm gw}) \leq 10^{-2} ~(10^{-4})$ under Gaussian ($\chi^2$) assumptions.}. 

\begin{table}
\centering
\begin{tabular}{||c| c |} 
 \hline
 Parameters $\boldsymbol{\theta}_{\rm gw}$ & Priors  \\ 
 \hline\hline
 $N_{\rm CMB}$  & U [$50,60$] \\
 \hline
 $f_{0}$ & logU [$10^{-6},~10$] \\
 \hline
 $\phi_{\rm end}$ &  U [0,~25] \\
 \hline
 $A_{+}$ & logU [$10^{-20},~10^{-6}$] \\
 \hline
 $A_{-}$ & logU [$10^{-20},~10^{-6}$] \\
 \hline
 $V_{0}$ & logU [$10^{-20},~10^{-6}$] \\
 \hline
 $\alpha_{\rm f}$  & U [$0, 250$] \\
 \hline
 $g$ & logU [$10^{-5}, ~1$] \\
 \hline
\end{tabular}
\caption{Priors applied to the Bayesian inference search, where we use uniform ($\rm U$) and log-uniform ($\rm logU$) distributions.
}
\label{table:priors}
\end{table}

\subsubsection{O1-O3 LVK data}
\label{sec: PE_O3b}

We now discuss the results of the Bayesian inference search using O1-O3 LVK data~\cite{LIGOScientific:2019lzm,KAGRA:2023pio}. 
We perform this analysis assuming a GWB sourced by both a background by Compact Binary coalescences (CBCs) and the toy model, from now on referred to as the TM + CBC model. 
We use priors on the toy model parameters, given in Table~\ref{table:priors}. In the frequency range of interest, the CBC background is well-described by an $f^{2/3}$ power law~\cite{PhysRevX.6.031018}. The parameter $\Omega_{\rm ref}$ that expresses the amplitude of the CBC background
amplitude at $25$ Hz, is included in the Bayesian inference search, with priors of LogU$(10^{-13}, 10^{-5})$. These priors are 
motivated by current estimates~\cite{Abbott:2017xzg, PhysRevD.104.022004}, which indicate its significance as a component of any GWB signal.

\begin{figure*}
    \centering
    \vspace{-5mm}
    \includegraphics[width=\textwidth]{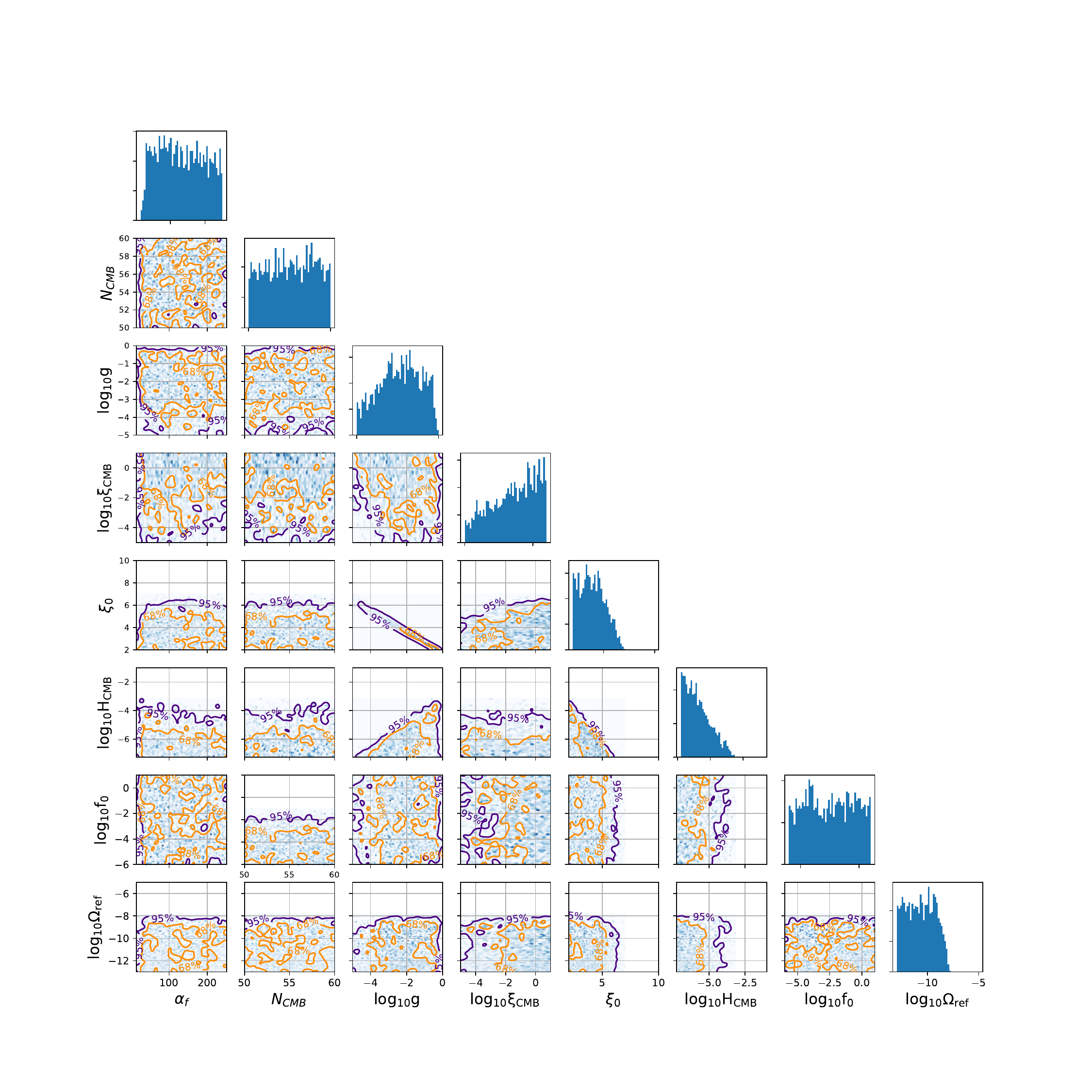}
    \vspace{-16mm}
    \caption{ Posterior distributions from Bayesian inference assuming a TM+CBC model, with $\chi^2$ statistics. Contour regions in yellow correspond to $68 \%$ CL and those in purple to $95 \%$ CL.
    In the upper part of the corner plot, the 1-dimensional marginalised posterior distribution is shown.}
    \label{fig:PE_O3_ChiSq}
\end{figure*}

\begin{figure*}
    \centering
    \vspace{-5mm}
    \includegraphics[width=\textwidth]{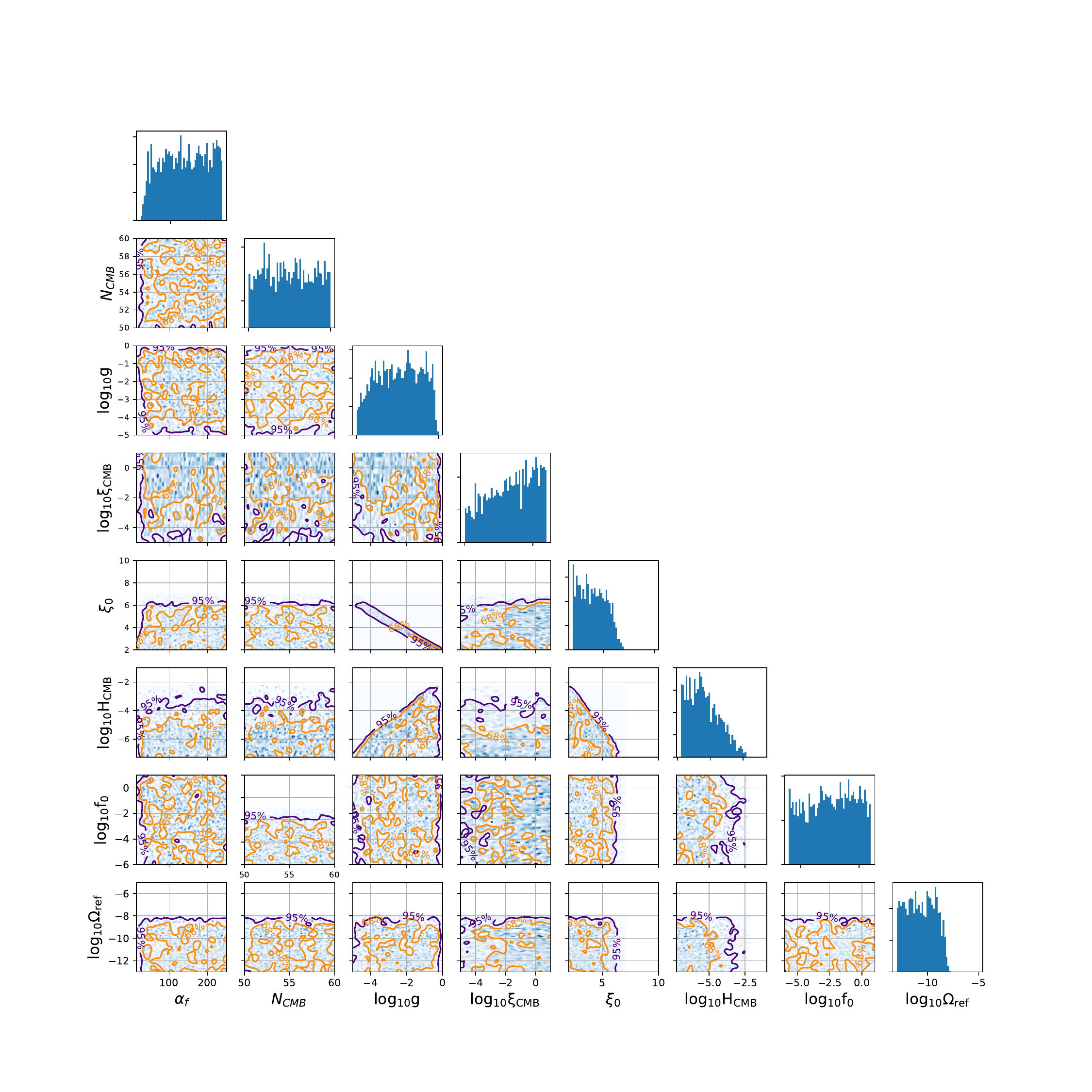}
    \vspace{-16mm}
    \caption{Posterior distributions from Bayesian inference assuming a TM+CBC model, with Gaussian statistics. Contour regions in yellow correspond to $68 \%$ CL and those in purple to $95 \%$ CL.
    In the upper part of the corner plot, the 1-dimensional marginalised posterior distribution is shown. }
    \label{fig:PE_O3_Gauss}
\end{figure*}

In Figs.~\ref{fig:PE_O3_ChiSq} and \ref{fig:PE_O3_Gauss}, we show the resulting posterior distributions, employing $\chi^2$ and Gaussian statistics for the PBH bound, respectively, obtained from a Bayesian inference search. 
Furthermore, the 2D contours for the different parameters $\boldsymbol{\theta}_{\rm gw}$ are displayed, highlighting those corresponding to the $68 \%$ CL in yellow, and those to the $68 \%$ CL in purple. 
Furthermore,
the posterior distributions allow us to establish $95\%$ CL limits on different parameters, given in Table~\ref{table:UL_O3}. For example, the UL on $\Omega_{\rm ref}$ is comparable to the previously reported UL on $\Omega_{\rm ref}(25 \text{ Hz})$ of $3.4 \times 10^{-9}$ ~\cite{KAGRA:2021kbb}.

\begin{table}
\centering
\begin{tabular}{||c| c| c |} 
 \hline
 Parameters %$\boldsymbol{\theta}_{\rm gw}$ 
 & $95\%$ CL in the Gaussian case & $95\%$ CL in the $\chi^2$ case  \\ 
 \hline\hline
 $\Omega_{\rm ref}$  &\rm $< 3.32 \times 10^{-9}$ &\rm $< 3.20 \times 10^{-9}$\\ 
 \hline
 $g$ & \rm $< 0.383$  & \rm $< 0.365$ \\
 \hline
 $V_0$ & \rm $< 1.72 \times 10^{-7}$ & \rm $< 4.21 \times 10^{-9}$ \\
 \hline
 $A_+$ & \rm $< 4.90 \times 10^{-11}$ & \rm $< 1.46 \times 10^{-12}$ \\
 \hline
 $A_-$ & \rm $< 8.69 \times 10^{-9}$ & \rm $< 1.94 \times 10^{-10}$ \\
 \hline
 $\xi_0$  & \rm $< 6.025$ & \rm $< 6.018$\\ 
 \hline
 $H_{\rm CMB}$  & \rm $< 2.47 \times 10^{-4}$ & \rm $< 3.87 \times 10^{-5}$\\ 
 \hline
\end{tabular}
\caption{$95\%$ CL on %some of 
the model's parameters obtained assuming Gaussian and $\chi^2$ statistics for the PBH bound}. 
\label{table:UL_O3}
\end{table}

Upper limits on $\xi_0$ and $H_{\rm{CMB}}$ can be formed from the non-observation of GWs modeled by Eqs.~\eqref{eq: GW_SU2}-~\eqref{eq: GW_vac}. Small and large combinations of $g$ and $\xi_0$ were ruled out; the smaller combinations falling within a non-Abelian regime and the larger combinations falling within a large backreaction regime. These constraints fit into the predictions in Fig.~\ref{fig: GeneralParam_H_range}.  One cannot place constraints on $N_{\rm{CMB}}$, $\phi_{\rm{end}}$, $\xi_{\rm{CMB}}$ and $f_0$ as to do so one needs GW data at smaller frequencies or, equivalently,  higher network sensitivity.

We also compute the Bayes factors comparing the hypothesis of data containing a TM+CBC signal versus noise only. The Bayes factor for the Gaussian statistic is 
$\ln(\mathcal{B}_{\rm Noise}^{\rm TM+CBC}) = -0.73 $, while that in the $\chi^2$ statistic is equal to $\ln(\mathcal{B}_{\rm Noise}^{\rm TM+CBC}) = -0.55$, both indicating no evidence for a TM+CBC signal in the data.

In conclusion, there is no evidence in the O1-O3 LVK data for a CBC background nor for a signal coming from the toy model. However, we can set $95\%$ CL ULs on some of the model's parameters. Let us also note that there is no noticeable difference between the two different statistics used.

\subsubsection{Future prospects}
\label{sec: PE_prospects}
We now aim to understand the prospects of detection and constraining SU(2) axion models with future ground-based detectors.
We perform a Bayesian inference search by using mock data consisting of the expected noise of the future detector's sensitivity, where we inject a GWB arising from the Toy Model that is the same as the one
plotted in orange in Fig.~\ref{fig: Model_GW_examples}.
It is worth highlighting that this example fits within current CMB bounds and sources a $\Delta_s^2 \leq 10^{-4}$.

We simulate a GWB search with the Advanced LVK sensitivity A+, an ET detector, a combined ET and CE network, and an ET + 2 CE network. 
The ET detector is assumed to be in the triangular configuration with assumed ET-D sensitivity~\cite{Hild:2010id} at the current Virgo detector site, the first CE is placed at the current Hanford site and the second CE is placed at the current Livingston site. Both CE's are assumed to have the sensitivity derived in~\cite{LIGOScientific:2016wof}. Similar to the previous searches, we assume Gaussian and $\chi^2$ statistics on the PBH bound in the searches. 

\begin{figure*}
    \centering
    \vspace{-5mm}
    \includegraphics[width=\textwidth]{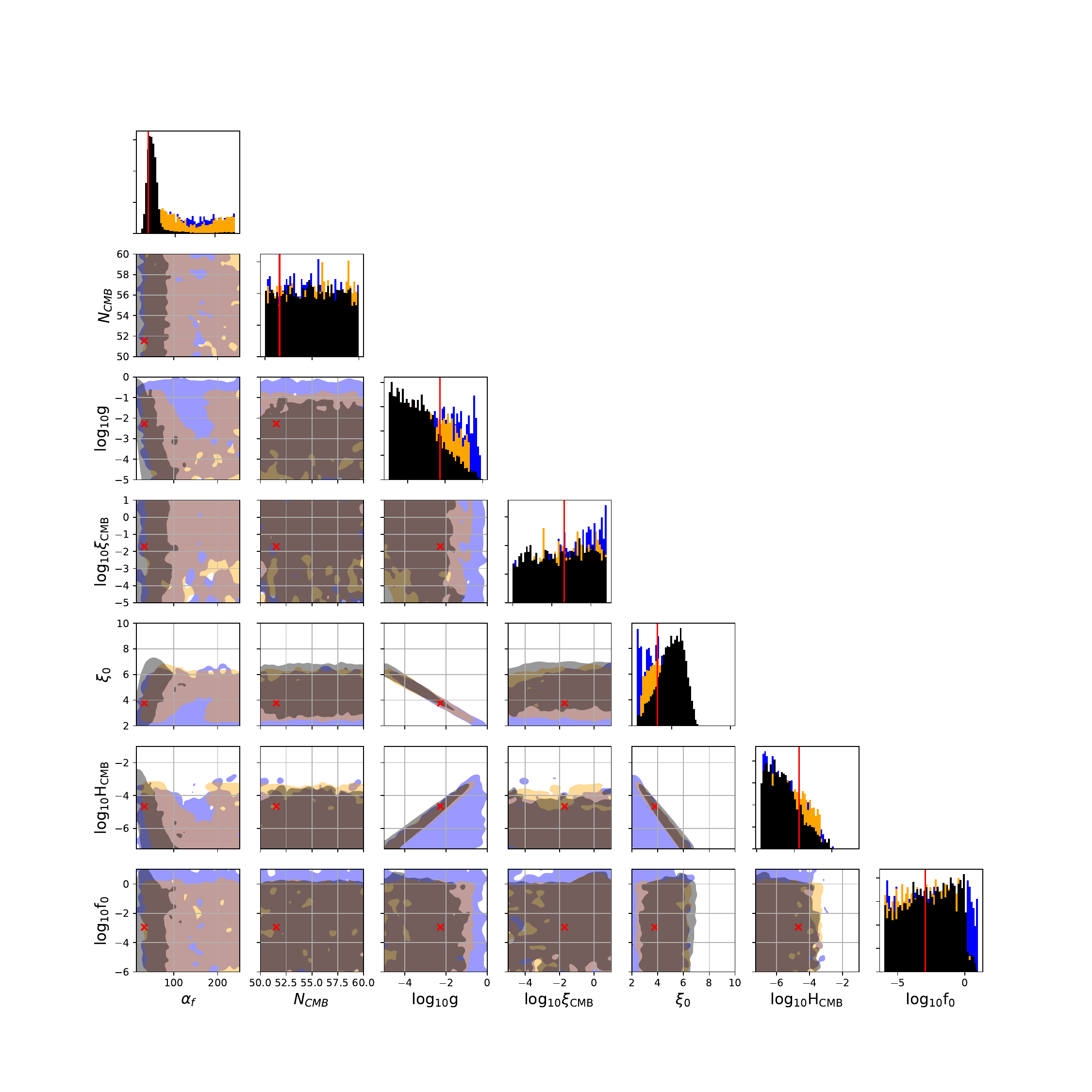}
    \vspace{-16mm}
    \caption{Posterior distributions from Bayesian inference assuming TM, with Gaussian statistics. The shaded regions correspond to the $95 \%$ CL.
    The different colors used are: blue for A+, orange for ET, and black for a network consisting of ET+2CE. The red cross and line correspond to the injected values.}
    \label{fig:PE_prospects_Gauss}
\end{figure*}

\begin{figure*}
    \centering
    \vspace{-5mm}
    \includegraphics[width=\textwidth]{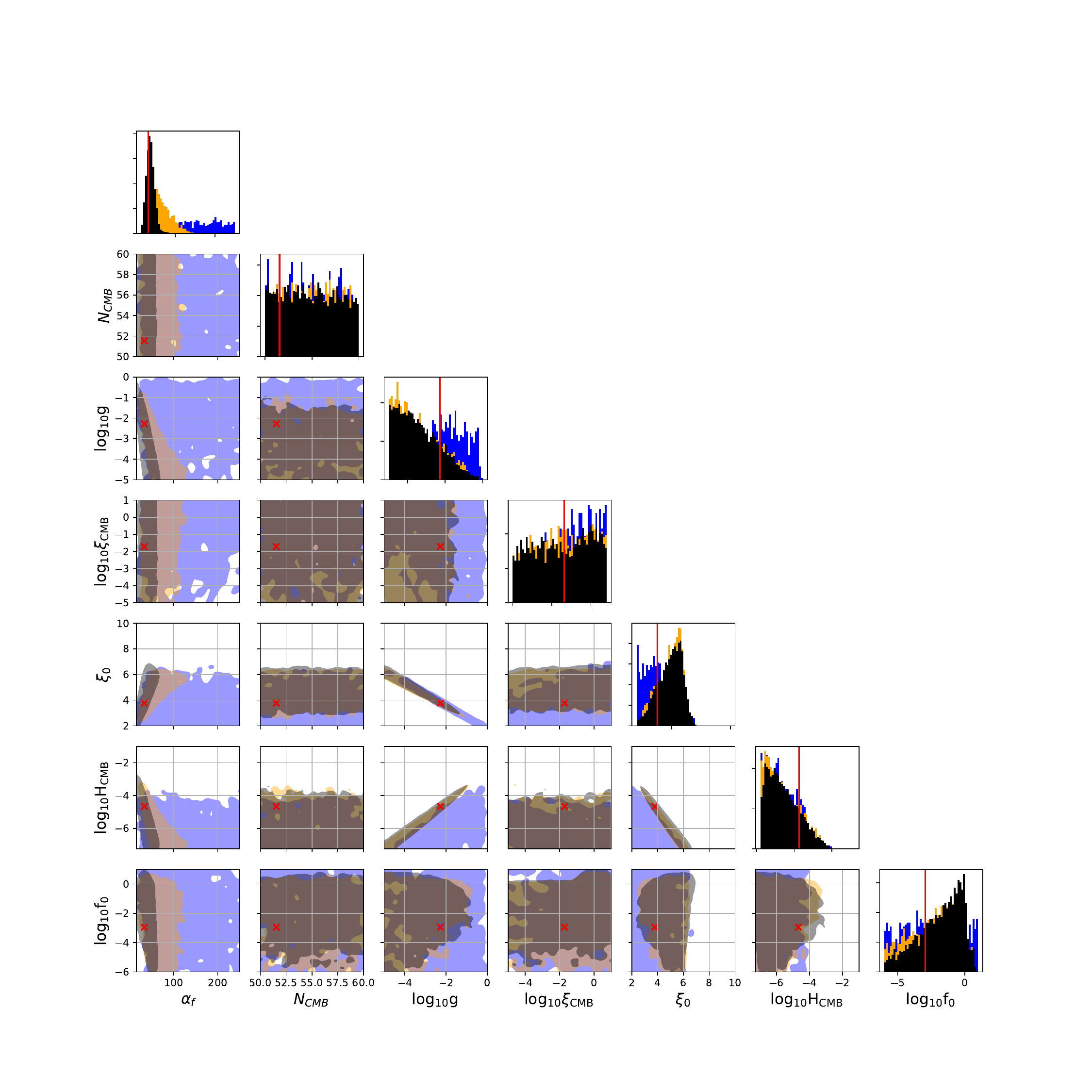}
    \vspace{-16mm}
    \caption{Posterior distributions from Bayesian inference assuming TM, with $\chi^2$ statistics. The shaded regions correspond to the $95 \%$ CL.
    The different colors used are: blue for A+, orange for ET, and black for a network consisting of ET+2CE. The red cross and line correspond to the injected values.}
    \label{fig:PE_prospects_ChiSq}
\end{figure*}

We show in Fig.~\ref{fig:PE_prospects_Gauss} the results for the Gaussian statistics, and in Fig.~\ref{fig:PE_prospects_ChiSq} for the $\chi^2$ one. 
%\CB{Talk about how constraints are placed or tightened.} 
It is easy to see that the addition of 3g detectors dramatically improves parameter constraints. Most notably, 3g networks are able to place upper bounds on coupling constants $\alpha_f$ and $g$, and a lower bounds on $\xi_0$ due to measurable presence of a GWB.
A constraint on $\alpha_f$ could be formulated under $\chi^2$ assumptions over Gaussian due to the more restrictive allowed physical space. Interestingly, the constraints from an ET search and those from an ET+CE search are effectively the same. This is because the measured cross-correlated strain in ET-CE combinations measured mainly detector noise.
With this, only LVK A+, ET, and ET+2CE search results are plotted. The retrieved logarithmic Bayes factors of each search can be seen in Table~\ref{table:Bayes_Prospects}. The 2D CL are the best constrained for an ET + 2CE network and the largest for LVK A+. Confident detections can be made with at least one ET present in the network, with $\log\mathcal{B}_{\rm{noise}}^{\rm{TM}} > 59$. %\footnote{We noted a Bayesian preference for Gaussian statistics assumptions over $\chi^2$ assumptions. This is believed to be because $\chi^2$-statistics assumptions confines the prior sampling space more than the Gaussian case, thus affecting the Bayesian posterior results.}

%\HD{should we comment that the Gaussian statistic is slightly preferred over the $\chi^2$ one? Maybe that could be interesting?} \CB{Definitely - I noticed the same thing but wasn't sure why quite yet...}

\begin{table}
\centering
\begin{tabular}{||c| c| c |} 
 \hline
 Detector Network & Gaussian case & $\chi^2$ case  \\ 
 \hline\hline
 LVK A+ &\rm -0.298 &\rm -0.132\\ 
 \hline
 ET  & \rm 61.07 & \rm 59.81 \\
 \hline
 ET + 2 CE & \rm 80.97  & \rm 81.15 \\
 \hline
\end{tabular}
\caption{Bayes factors $\ln\mathcal{B}_{\rm{noise}}^{\rm{TM}}$ of parameter estimation for different detector networks and for both Gaussian and $\chi^2$ PBH statistic assumptions. }
\label{table:Bayes_Prospects}
\end{table}

\section{Conclusions}
\label{sec: conclusions}
In the framework of axion-SU(2) gauge inflation, we have investigated  GWB detectability with current and third-generation interferometers.
In our analysis we have included previously unconsidered astrophysical and cosmological constraints, namely those from CMB measurements as well as PBH overproduction.
We have studied the characteristics of the inflationary potential that would lead to an inflationary velocity 
capable of evading
such constraints while producing a potentially detectable GWB signal. 
The observationally allowed parameter space of well-studied inflationary potentials with fast growing inflaton velocity within the framework of SU(2) gauge inflation, 
appears
to lead to a GWB undetectable by current or 3g interferometers.
We have explicitly shown that once CMB and PBH constraints are imposed to the parameters of Starobinsky inflation, the GWB remains effectively undetectable with a 3g detector network, like ET+2CE.

However, models like a piecewise toy model can conform to physical and observational considerations and can be detectable with 3g detectors. Analysing an injection with at least one ET in a triangular configuration would yield $\log\mathcal{B}_{\rm{noise}}^{\rm{TM}} > 59$. Introducing additional CEs improves both CL regions and detection prospects. Although such a model is unlikely to be detected in LVK, one can still place constraints based on the non-observation of physically allowed spectra. Parameter estimation searches for a combined Toy Model and CBC background model on O1-O3 data under assumed Gaussian and $\chi^2$ statistics for their PBH bound indicated no evidence for a combined model in GW data, but were able to place parameter space bounds. %Since other potentials are expected to result in growing inflaton velocity, 
Since other inflaton potentials predict growing inflaton velocity that results in a larger GWB amplitude,
the analysis conducted assuming the non-evolving toy model places conservative constraints in the ground detector frequency range.

More broadly, we explore general GW detection prospects using conservative assumptions on inflaton velocity behavior and PBH bounds. We find that working, detectable models are limited to a narrow parameter space, requiring idealised conditions such as approximately constant inflaton velocities in the ground detector frequency regime.

This methodology can be applied to other models outside of the two explored in this work, in order to make predictions on such model's detection prospects in present and future ground detector networks. One could also use this general approach in model development to build detectable inflationary models compatible with theoretical and observational constraints. Furthermore, a better theoretical understanding of SU(2) mechanisms in the strong backreaction regime can help project a complete picture of ground detector detection prospects. 

\begin{acknowledgments}
We thank Alberto Mariotti and Alex Sevrin, who were involved in the early stages of this study.

We acknowledge computational resources provided by the LIGO Laboratory and supported by National Science Foundation Grants No. PHY-0757058 and No. PHY-0823459. This doccument has LIGO DCC number LIGO-P2400223. The work of MS is partially supported by the Science and Technology Facilities Council (STFC grant ST/X000753/1).
ARR is supported in part by the Strategic Research Program "High-Energy Physics" of the Research Council of the Vrije Universiteit Brussel and by the iBOF "Unlocking the Dark Universe with Gravitational Wave Observations: from Quantum Optics to Quantum Gravity" of the Vlaamse Interuniversitaire Raad and by the FWO IRI grant I002123N "Essential Technologies for the Einstein Telescope".
SK is supported by the Spanish Atracci\'on de Talento contract no. 2019-T1/TIC-13177 granted by Comunidad de Madrid, the Spanish Research Agency (Agencia Estatal de Investigaci\'on) through the Grant IFT Centro de Excelencia Severo Ochoa No CEX2020-001007-S, the I+D grant PID2020-118159GA-C42, and the Consolidaci\'on Investigadora 2022 grant CNS2022-135211, funded by MCIN/AEI/10.13039/501100011033, and Japan Society for the Promotion of Science (JSPS) KAKENHI Grant no. JP20H05853, and JP23H00110, JP24K00624.

This research has made use of data or software obtained from the Gravitational Wave Open Science Center (gwosc.org), a service of the LIGO Scientific Collaboration, the Virgo Collaboration, and KAGRA. This material is based upon work supported by NSF's LIGO Laboratory which is a major facility fully funded by the National Science Foundation, as well as the Science and Technology Facilities Council (STFC) of the United Kingdom, the Max-Planck-Society (MPS), and the State of Niedersachsen/Germany for support of the construction of Advanced LIGO and construction and operation of the GEO600 detector. Additional support for Advanced LIGO was provided by the Australian Research Council. Virgo is funded, through the European Gravitational Observatory (EGO), by the French Centre National de Recherche Scientifique (CNRS), the Italian Istituto Nazionale di Fisica Nucleare (INFN) and the Dutch Nikhef, with contributions by institutions from Belgium, Germany, Greece, Hungary, Ireland, Japan, Monaco, Poland, Portugal, Spain. KAGRA is supported by Ministry of Education, Culture, Sports, Science and Technology (MEXT), Japan Society for the Promotion of Science (JSPS) in Japan; National Research Foundation (NRF) and Ministry of Science and ICT (MSIT) in Korea; Academia Sinica (AS) and National Science and Technology Council (NSTC) in Taiwan.

\end{acknowledgments}

\newpage

\bibliography{AxInfbib}

\end{document}